\documentstyle[sprocl,psfig]{article}

\def\be{\begin{equation}}
\def\ee{\end{equation}}
\def\bea{\begin{eqnarray}}
\def\eea{\end{eqnarray}}
\def\bean{\begin{eqnarray*}}
\def\eean{\end{eqnarray*}}
\def\half{\frac{1}{2}}
\def\vecnul{{\bf 0}}
\def\vecp{{\bf p}}
\def\veck{{\bf k}}
\def\vecx{{\bf x}}
\def\vecv{{\bf v}}
\def\vecE{{\bf E}}

\def\ep{\epsilon}
\def\lm{\lambda}
\def\Lm{\Lambda}
\def\om{\omega}
\def\Om{\Omega}
\def\bra{\langle}
\def\ket{\rangle}

\begin{document}

\title{RENORMALIZABILITY OF HOT CLASSICAL FIELD THEORY}

\author{GERT AARTS}

\address{
Institute for Theoretical Physics, Utrecht University\\
Princetonplein 5, 3508 TA Utrecht, the Netherlands}

\maketitle\abstracts{
I discuss the possibility of using classical field theory to approximate 
hot, real-time quantum field theory. I calculate, in a scalar theory, the 
classical two point and four point function in perturbation theory. 
The counterterms needed to make the classical correlation functions finite 
are dictated by the superrenormalizability of the static theory. The 
classical expressions approximate the quantum ones, when the classical 
parameters are chosen according to the dimensional reduction matching 
rules.  I end with an outlook to gauge theories.}
  
\section{Introduction}
The classical approximation was suggested several years ago \cite{GrRu} 
to make non-perturbative calculations of real-time correlation functions 
at finite temperature possible, namely by the use of numerical simulations 
\cite{Kretc}.
In this talk I discuss whether classical field theory can be used to 
approximate the quantum theory, by doing a perturbative calculation in 
$\lm\phi^4$ theory. Almost everything I say can be found in ref 
\cite{AaSm}, written together with Jan Smit.

\section{Dimensional reduction for static quantities}
\label{section2}
I start with a brief introduction to the dimensional 
reduction approach for time independent quantities \cite{Kaetal}.
The question is how to calculate (in a reliable way) static equilibrium 
quantities, 
such as the critical temperature and the order of the phase transition, in a 
weakly coupled 
quantum field theory at finite temperature $T$. 
In the Matsubara formalism, the field $\phi(\vecx, \tau)$ obeys periodic 
boundary conditions in imaginary time, $\phi(\vecx, 0) = \phi(\vecx, 
\beta)$, and hence can be decomposed as
\[ \phi(\vecx, \tau) = T \sum_{n} e^{i\omega_n\tau}\phi_n(\vecx),\]
with the Matsubara frequencies $\omega_n = 2\pi nT$.
The dimensional reduction approach is the following: an effective 3-d 
theory is constructed for $\phi_0(\vecx)$, which can be used e.g. 
non-perturbatively. This effective theory is 
chosen to have the same form as the $\phi_0$ part of the original theory, 
but with effective parameters. The parameters in the 3-d theory, 
$m^2_{\rm eff}$ and $\lm_{\rm eff}$,  are 
determined by perturbatively matching 3-d correlation functions to the 
full 4-d correlation functions. A nice property of the effective 
theory is that it is superrenormalizable. Hence, 
the coupling constant only receives a finite correction due to matching, 
$\lm_{\rm eff} \approx \lm$.   
The effective mass parameter is written as $m^2_{\rm eff} = m^2 -\delta 
m^2$. $\delta m^2$ takes care of the divergences in the 3-d theory, 
namely in the one loop tadpole and the two loop setting sun diagram. The 
finite part $m^2$ contains in particular the thermal mass, $m^2 \approx 
\lm T^2/24\hbar$, where I indicated explicitly the $\hbar$ dependence.

\section{What about time dependent correlation functions?}
The question is now if this approach can be  extended 
to  calculate also  real-time quantities, such as 
transition rates, the plasmon frequency and damping rates.

Consider the partition function for the effective theory ($\phi = \phi_0$)
\[ Z_{\footnotesize\mbox{3-d}} = \int {\cal D}\phi\,e^{-\beta V(\phi)}, 
\;\;\;\;\;\;
V = \int d^3x\left( \half (\nabla \phi)^2 + \half m^2_{\rm eff}\phi^2 + 
\frac{\lm_{\rm eff}}{4!} \phi^4\right).\]
The form is that of the classical partition function,
\[ Z_{\rm cl} = \int {\cal D}\pi{\cal D}\phi\,e^{-\beta H(\pi, \phi)}, 
\;\;\;\;\;\;\;\;
H = \int d^3x\,\half \pi^2 + V(\phi),\]
after the integration over the momenta has been performed. Therefore, 
let's take a look at time dependent {\em classical} correlation functions.
The classical two point function is given by ($x = (\vecx, t)$)
\be
\label{eqS}
 S(x-x') = \bra \phi(x)\phi(x')\ket_{\rm cl} 
= Z_{\rm cl}^{-1} \int {\cal D}\pi{\cal D}\phi\,e^{-\beta H(\pi, \phi)}\,
\phi(x)\phi(x'),\ee
with $\phi(x)$ the solution of the classical equations of motion 
$\dot{\phi}(x) = \{\phi(x), H\}$, $\dot{\pi}(x) = \{\pi(x), H\}$,
with the initial conditions $\phi(\vecx, t_0) = \phi(\vecx)$, $\pi(\vecx, 
t_0) = \pi(\vecx)$. The integration over phase space is over the initial 
conditions at  $t=t_0$, weighted with the Boltzmann weight.

This definition (\ref{eqS}) raises two obvious questions
\begin{itemize}
\item What about the divergences? When $t=t'$, the integration over the 
momenta is trivial, and we are back in the 3-d superrenormalizable 
theory, what happens to the 3-d divergences when $t \neq t'$?
\item What about the relation with time dependent correlation functions 
in the full quantum field theory? Again, when $t=t'$, the matching 
relations make sure that the 3-d correlation functions approximate 
the full 4-d ones, what happens when $t \neq t'$?
\end{itemize}

\section{Perturbative expansion in the classical field theory}
To answer the above stated questions, I do a perturbative calculation of  
the two point and the four point function to order $\lm_{\rm eff}^2$. The 
perturbative  solution of the equations of motion takes the form 
\bean
\phi(x) &=& \int \frac{d^3k}{(2\pi)^3}\,e^{i\veck\cdot\vecx}\,[
\phi(\veck)\cos \omega_\veck(t-t_0) +
\frac{\pi(\veck)}{\omega_\veck}\sin \omega_\veck(t-t_0)
]\\
&& -\lm_{\rm eff} \int d^4x'\, 
G^R_0(x-x')\frac{1}{3!}\phi_0^3(x') + \ldots
\eean
Here $\omega_\veck^2 = \veck^2+m^2$, and the retarded Green function is 
defined by
\be
\label{eqGR}
 G^R_{\rm cl}(x-x') =  -\theta(t-t')\bra 
\{\phi(x),\phi(x')\}\ket_{\rm cl}.\ee
The free two point and retarded Green function are given by 
\renewcommand{\theequation}{\arabic{equation}a,b}
\be
\label{eqS0} S_0(\veck, t) =   T\frac{\cos \om_\veck t}{\om_\veck^2}, 
\;\;\;\;\;\;\;\;\;\;\;\;\;\; 
G^R_0(\veck, t) = \theta(t)\frac{\sin \om_\veck 
t}{\om_\veck}.\ee
It is now  straightforward to calculate the two point function, 
defined in (\ref{eqS}). This gives the connected two point function, but 
it is easy to 
identify the classical 1PI diagrams and in particular the retarded self 
energy.
It is convenient to work in (temporal) momentum space, just 
as in thermal field theory.
\renewcommand{\theequation}{\arabic{equation}}

Let me make a small sidestep here to show the relation with the quantum 
theory. The quantum retarded Green function is defined by
\[ G^R_{\rm qm}(x-x') = i\theta(t-t')\bra[\phi(x),\phi(x')]\ket_{\rm qm}.\]
The free retarded Green function is given by (3b), the classical 
expression. It is independent of $T$ and $\hbar$. The 
quantum analogon of the two point function is
\[ F(x-x') =  \half\bra \phi(x)\phi(x')+\phi(x')\phi(x)\ket_{\rm qm}.\]
In the free case it is
\[ 
F_0(\veck, t) = [n(\om_\veck)+\half]\frac{\cos \om_\veck t}{\om_\veck} =
T\sum_n \frac{\cos \om_\veck t}{\om_n^2+\om_\veck^2},
\;\;\;\;
n(\om) = \left(e^{\beta \om}-1\right)^{-1}.
\]
The classical two point function (3a) is just the $n=0$ term, similar  
as in the dimensional reduction approach in imaginary time.
The relation between diagrams in the classical and the quantum theory 
can be completely understood by a suitable formulation of the real-time 
formalism of thermal field theory.

\section{Classical retarded self energy and vertex function}

\begin{minipage}{5.4cm}
\vspace{0.1cm}
\hspace{0.3cm}
\psfig{figure=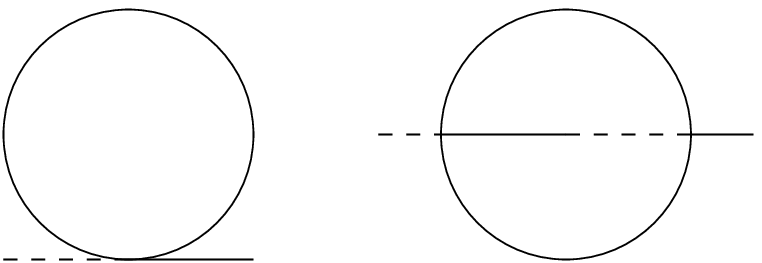,width=4.5cm}
\vspace{0.3cm}

{\footnotesize Figure 1: Retarded self energy: one loop and setting sun 
diagram. 
The full line represents $S_0$ and the dashed-full line $G_0^R$.}
\end{minipage}
 \ \hspace{0.5cm}  \
\begin{minipage}{5.5cm}
The one loop classical retarded self energy is given by the tadpole 
diagram in fig 1. It is independent of the external momenta, 
and hence specific problems related to time dependence  
play no role. The diagram is linear divergent, and the divergence is 
canceled with $\delta m^2$. The result is the same as in the dimensional 
reduction approach. 
\end{minipage}

\vspace{0.14cm}
The two loop setting sun diagram is more interesting. It has a 
complicated momentum dependence, a logarithmic divergence when $t=t'$, 
and it contains an imaginary part, which gives e.g. Landau damping.
The diagram (see fig 1) is
\bea
\nonumber
\Sigma_{R, \rm cl}^{\rm sun}(p) =
-\half \lm_{\rm eff}^2
\int\frac{d^4k_1}{(2\pi)^4}\frac{d^4k_2}{(2\pi)^4}
S_0(k_1)S_0(k_2)G^R_0(p-k_1-k_2)&&\\
=
\label{eqsun}
-\frac{\lm_{\rm eff}^2T^2}{6} 
\int \frac{d^3k_1}{(2\pi)^3}\frac{d^3k_2}{(2\pi)^3}
\frac{1}{\om_{\veck_1}^2 \om_{\veck_2}^2 \om_{\veck_3}^2}
+\frac{p^0}{T}\int \frac{d\Om}{2\pi}\frac{w(\vecp,\Om)}{p^0+i\ep +\Om},&&
\eea
with $\om_{\veck_3} = \om_{\vecp-\veck_1-\veck_2}$ and the typical 
scattering integral
\[ w(\vecp, \Om) =
\frac{\lm^2_{\rm eff}}{6}
\sum_{\{\pm\}}\int
\prod_{j=1}^3\left[ 
\frac{d^3k_j}{(2\pi)^32\om_{\veck_j}}\frac{T}{\om_{\veck_j}}\right]
(2\pi)^{4}
\delta(\Om \pm\om_{\veck_1}\pm\om_{\veck_2}\pm\om_{\veck_3}).
\]
$T/\om_\veck$ is the classical thermal distribution function, and the sum is 
over all $+$'s and $-$'s.
The first term in (\ref{eqsun}) is independent of $p^0$, real and is the 
standard dimensional reduction setting sun result. Its 
logarithmic divergence is canceled by $\delta m^2$.
The second term in (\ref{eqsun}) contains all the $p^0$ dependence, it is 
finite, it has an imaginary part and it represents the corresponding leading 
order quantum expression for large $T$ and small $\lambda_{\rm eff}$.
For example, the classical plasmon damping rate is
\be \label{eqdamping}
 \gamma_{\rm cl} =  -\frac{\mbox{Im}\; \Sigma^{\rm 
sun}_{R, \rm cl} (\vecnul, m)}{2m} = \frac{\lambda_{\rm eff}^2 T^2}{1536
\pi m} \approx \frac{\lambda\sqrt{\lm\hbar}T}{128\sqrt{6}\pi},
\ee
where I used the matching results from section \ref{section2}.  Note that 
$\hbar$ only enters through $m$.
(\ref{eqdamping}) is indeed the leading order quantum result
\[  \gamma_{\rm qm} = 
\frac{\lambda\sqrt{\lm\hbar}T}{128\sqrt{6}\pi}\left(1+ 
{\cal O}(\sqrt{\hbar\lm}\log \hbar\lm)\right).\]
Finally, the classical four point function is finite, as in the 
static superrenormalizable theory, 
and it reproduces the leading order quantum vertex function.

\section{Conclusion and outlook to gauge theories}
To summarize, I have considered $\lm\phi^4$ theory (to two loops), with the 
conclusion that the time dependent classical theory is renormalizable, 
i.e. local counterterms are sufficient to make the correlation functions 
finite. Furthermore the classical expressions approximate the quantum 
ones. Hard thermal loop effects (which is simply the thermal mass) are 
easily incorporated because of momentum independence. I stress that the 
strategy followed is not that of integrating out high momentum modes, 
which leads to an effective theory with an intermediate cutoff \cite{BoMcSm}.

Is there any use for gauge theories? Here the hard thermal loops effects  
are much more complicated because of  momentum dependence. The one 
loop classical self energy is linear divergent. The question is if this 
divergence can be canceled with a (non-local) counterterm. A way to do 
this might be to use the effective equations of motion, derived within a   
kinetic theory framework \cite{BlIa}, namely 
\bean
D_\mu F^{\mu\nu} = j^\nu,&&\;\;\;\;\;\;
v^\mu D_\mu w(x, \vecv) = \vecv\cdot \vecE(x),\\
j^{\mu}(x) = 2m^2\int \frac{d\vecv}{4\pi} v^\mu w(x, \vecv),&&\;\;\;\;\;\; 
m^2 = \frac{1}{6}g^2NT^2 - \frac{1}{\pi^2}g^2NT\Lm,
\eean
with $v^\mu = (1, \vecv)$, $\vecv\cdot\vecv =1$.
The mass parameter now also contains a momentum  cutoff ($\Lm$) dependent 
counterterm, just as in the scalar theory. To preserve gauge invariance, 
it should be formulated on a lattice.

\section*{Acknowledgments}
This work is supported by FOM.

\end{document}